\documentclass [onecolumn,glov2,superscriptaddress,final]{svjour2}   
\smartqed  
\usepackage{graphicx}
\journalname{Acta Physica Polonica A}
\begin{document}

\title{Different quantization mechanisms in single-electron pumps driven by
surface acoustic waves}

\titlerunning{Different quantization mechanisms ...}

\author{P.~Utko,$^1$ K.~Gloos,$^2$ J.~Bindslev Hansen,$^3$
    C.~B.~S{\o}rensen,$^1$ and P.~E.~Lindelof $^1$}

\authorrunning{P.~Utko {\it et al.}}

\institute{$^1$Nano-Science Center, Niels Bohr Institute, University
of Copenhagen, Universitetsparken 5, DK-2100 Copenhagen,
Denmark\\
$^2$Wihuri Physical Laboratory, Department of Physics, University of
Turku, FIN-20014 Turku, Finland\\
$^3$Department of Physics, Technical University of Denmark, DK-2800
Lyngby, Denmark}

\date{} 

\maketitle

\begin{abstract} {
We have studied the acoustoelectric current in single-electron pumps driven by
surface acoustic waves. We have found that in certain parameter ranges two
different sets of quantized steps dominate the acoustoelectric current versus
gate-voltage characteristics. In some cases, both types of quantized steps
appear simultaneously though at different current values, as if they were
superposed on each other. This could indicate two independent quantization
mechanisms for the acoustoelectric current.}

\end{abstract}

\section{Introduction}
\label{intro}
During the last two decades single-electron devices have attracted a
lot of interest. In one class of such systems, surface acoustic
waves (SAWs) are used to transfer a discrete number of electrons
across a quantum point contact (QPC) operated in the closed-channel
regime
\cite{Shilton1996,Cunningham2000,Ebbecke2000,Fletcher2003,Utko2003,Gloos2004}.
This is often described using a model of "moving quantum dots"
\cite{Shilton1996} where each dot, formed by the dynamic SAW
potential and the static barrier of the QPC, captures electrons at
the QPC entrance. Coulomb repulsion between the trapped electrons
restricts their number inside the dot. As a result, the SAW-induced
acoustoelectric current develops plateaus at $I = nef$, where $n$ is
an integer, $e$ the electron charge, and $f$ the SAW frequency. The
current quantization can be observed as a function of the gate and
bias voltages, the SAW power, and frequency.

Here we have investigated the response of the acoustoelectric
current to changes in the gate voltage $V_g$ and the SAW frequency
$f$. Within a certain frequency range, two separate sets of plateaus
seemed to dominate the $I(V_g)$ characteristics. In some cases, both
types of quantized steps appeared simultaneously, though at
different current values, as if they were superposed on each other.
Their presence could result from two independent quantization
mechanisms for the acoustoelectric current.

\section{Experiment}
\label{experiment}

\begin{figure}
 \begin{center}
  \includegraphics[width=1\textwidth]{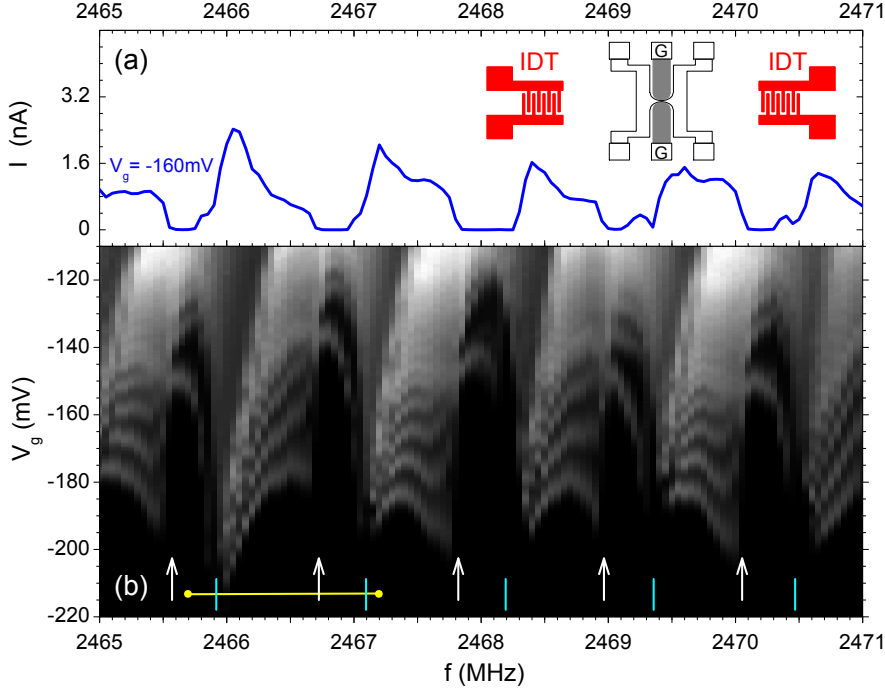}
 \end{center}
\caption{(a) Acoustoelectric current $I$ as a function of the SAW
frequency $f$. Inset shows a schematic layout of the device. (b)
Gray-scale coded plot of transconductance $dI/dV_g$ with respect to
both the SAW frequency $f$ and gate voltage $V_g$. Dark (light)
indicates small (large) values of the current derivative. The
vertical arrows and bars mark the frequencies around which one set
of acoustoelectric plateaus is replaced by another one, see text for
details. The horizontal line indicates the frequency range from
which the traces in Fig.~\ref{results-2C} are selected. All
measurements were taken at $P = +9.8\,$dBm and $T = 1.8\,$K.}
\label{fig1}
\end{figure}

The inset to Fig.~\ref{fig1}(a) shows a schematic layout of our
devices. Two aluminum interdigital transducers (IDTs) can be used to
generate the SAW. They are deposited 2.6\,mm apart, on both sides of
a 2DEG mesa with a QPC in the center. The IDT electrode spacing sets
the fundamental wavelength and frequency of the SAW to about
$1.15\,\mu$m and 2.45\,GHz, respectively. The GaAs/AlGaAs
heterostructure has a mobility of $105\,$m$^2/$(Vs) and a carrier
density of $2.8 \cdot 10^{15}$\,m$^{-2}$, measured in the dark at
10\,K. The QPC is patterned by electron-beam lithography. Two
semicircular trenches are shallow-etched to form a smooth
constriction between two electron reservoirs, whereas large areas of
the 2DEG across the channel serve as side gates. Depending on the
device, the trenches have a curvature radius of 5.0, 7.5 or
10.0\,$\mu$m. They are 200\,nm wide and 40\,nm deep. See
Refs.~\cite{Utko2003,Gloos2004} for more details on the device
layout.

The rf excitation of power $P$ could be applied to one of the two
IDTs, or split up and simultaneously fed to both transducers. With a
phase shifter and an attenuator in one of the rf lines, the relative
magnitude and the relative phase of both signals could then be
varied and adjusted. A low-noise current preamplifier detected the
acoustoelectric current. The samples were investigated either in a
$^3$He refrigerator operating at 1.2\,K or a $^4$He refrigerator
operating at 1.8\,K. The lower base temperature of the $^3$He
cryostat did not help to improve the results because in both systems
the 2DEG of our devices was heated up to around 5\,K at the
typically applied rf powers of around $10-15\,$dBm \cite{Utko2006}.

\begin{figure}
  \begin{center}
   \includegraphics[width=1\textwidth]{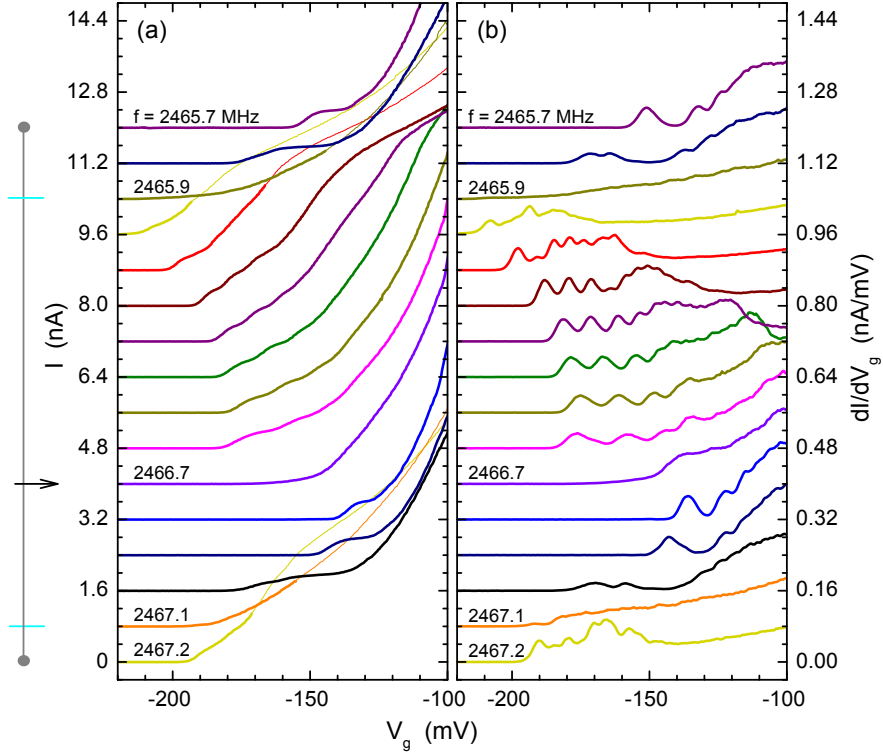}
  \end{center}
\caption{(a) Acoustoelectric current $I$ and (b) transconductance
$dI/dV_g$ as a function of gate voltage $V_g$. The 16 current
(transconductance) traces were recorded at fixed SAW frequencies
from 2465.7 to 2467.2\,MHz, in steps of 0.1\,MHz. The curves were
successively offset in the vertical direction by 0.8\,nA
(0.08\,nA/mV). The measurements were carried out at $P = +9.8\,$dBm,
and $T = 1.8\,$K.} \label{results-2C}
\end{figure}

Figures~\ref{fig1}(a) and ~\ref{results-2C}(a) show typical $I(f)$
and $I(V_g)$ characteristics, respectively, obtained when only one
IDT was used to generate the SAW. The pronounced 1.1\,MHz beat
period of the current indicates that the interference of the surface
acoustic wave with reflected waves matters. This is supported by the
results obtained when a second, independent SAW beam was added,
traveling in opposite direction \cite{Utko2006b}. Current plateaus
at integer multiples of $ef$ can be resolved in either $I(f)$ or
$I(V_g)$ traces. However, they become more apparent in the current
derivative, see Figs.~\ref{fig1}(b) and ~\ref{results-2C}(b). Thus,
to enhance the resolution of the acoustoelectric transitions, we
have routinely monitored the transconductance $dI/dV_g$ along with
the current $I$. This was done by adding a small 117\,Hz modulation
of $dV_g \approx 0.5\,$mV to the gate voltage, and measuring the ac
component of the current with a lock-in amplifier.

The gray-scale plot of $dI/dV_g$ in Fig.~\ref{fig1}(b) reveals a
periodic structure with respect to frequency, with a beat period of
roughly $1.1\,$MHz. However, transconductance minima corresponding
to current plateaus do not evolve smoothly over the entire 1.1\,MHz
period. On increasing the SAW frequency, broad plateaus with respect
to $V_g$ are abruptly replaced by densely-packed quantized steps.
Frequencies around which such transitions occur are marked with
vertical bars. Arrows indicate another type of transition where the
acoustoelectric current drastically shifts its onset along the
$V_g$-axis in response to a small change in frequency.

\begin{figure}
  \begin{center}
   \includegraphics[width=1\textwidth]{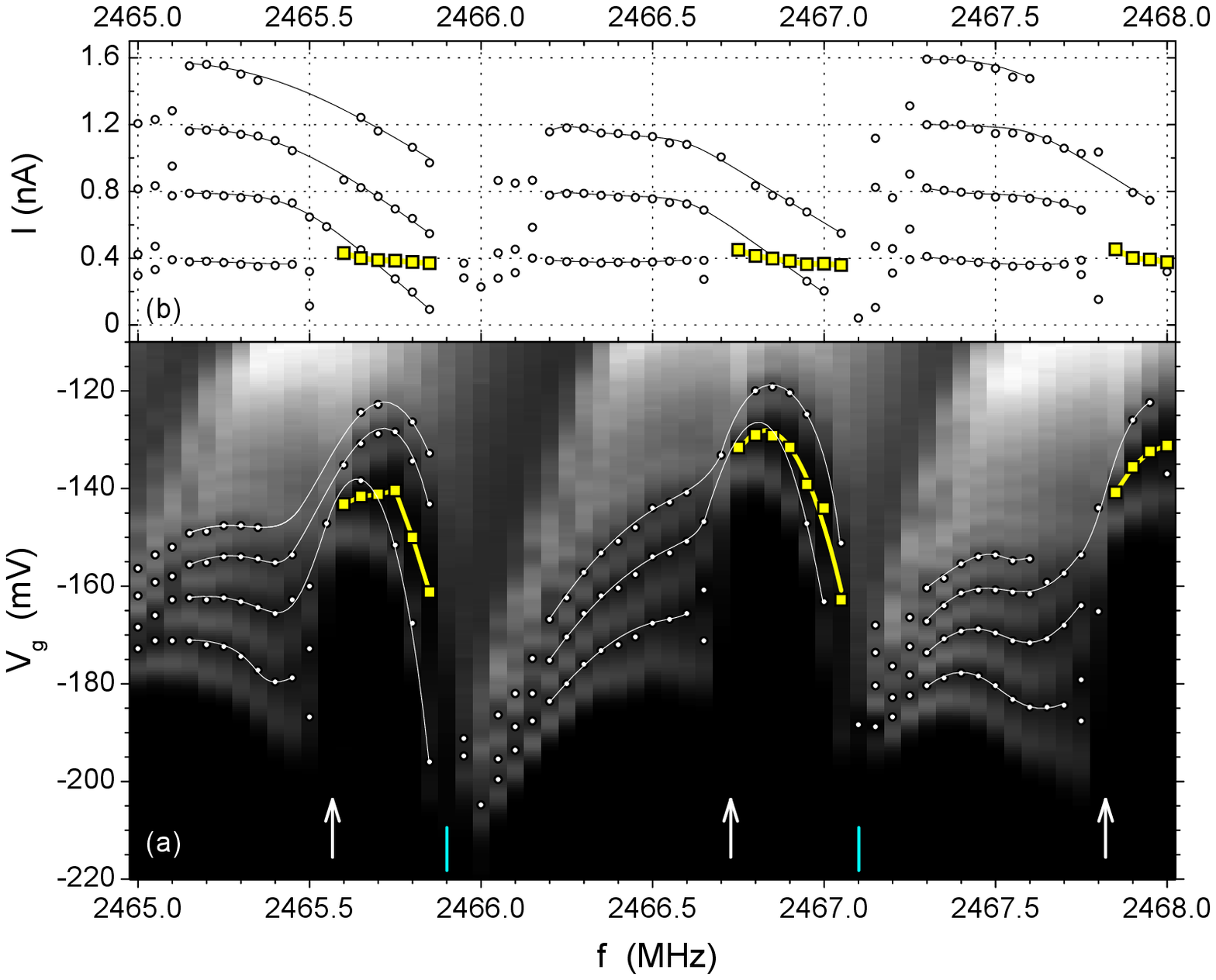}
  \end{center}
\caption{(a) Gray-scale plot of transconductance $dI/dV_g$ with
respect to SAW frequency $f$ and gate voltage $V_g$. Open circles
and full squares mark the position of transconductance minima. (b)
Acoustoelectric current $I$ at the minima indicated in (a). In both
(a) and (b), two different kinds of acoustoelectric plateaus can be
distinguished. Solid lines through the data points are guides to the
eye. The measurements were carried out at $P = +9.8\,$dBm and $T =
1.8\,$K.} \label{plateaus-2C}
\end{figure}

Those features become more apparent in Fig.~\ref{results-2C},
showing selected $I(V_g)$ and $dI(V_g)/dV_g$ traces taken at
frequencies from 2465.7 to 2467.2\,MHz. In both Fig.~\ref{fig1}(b)
and ~\ref{results-2C}, two frequency intervals can be distinguished
within a beat period of $\sim1.1\,$MHz where two different sets of
plateaus dominate the $I(V_g)$ characteristics. Those two sets seem
to replace each other around certain frequencies, indicated by
arrows and bars. However, in some cases both sets can also appear
simultaneously in the $I(V_g)$ traces, as if they were superposed
onto each other. The dominating set consists then of broad current
plateaus with respect to the gate voltage that are well-defined at
the expected multiples of $ef$. On the other hand, weakly-pronounced
steps belonging to the second set are formed below those ideal
values, as highlighted in Fig.~\ref{plateaus-2C}. Both sets respond
differently to the SAW frequency, as for example in the range from
2465.7 to 2465.9\,MHz in Fig.~\ref{plateaus-2C}. When the frequency
is incremented within such a range, one plateau in $I(V_g)$ remains
close to the expected value of $ef \approx 400\,$pA, while the other
set of quantized steps appears at lower and lower currents.

\section{Discussion and conclusions}
\label{discussion}

The frequency response measurements presented here reveal a
complicated pattern of the acoustoelectric transitions. Their most
apparent feature, the 1.1\,MHz beating of the acoustoelectric
current, immediately reminds us of the interference patterns due to
a standing wave. In fact, the results obtained while varying the SAW
frequency closely resemble those when the phase is controlled
between two counter-propagating SAW beams \cite{Utko2006b}. The
period of 1.1\,MHz thus corresponds to a phase shift of $2\pi$. This
already indicates that the standing wave matters, but it does not
specify where to find its nodes.

The SAW-driven single-electron transport is usually described using
a model of moving quantum dots \cite{Shilton1996}. In such a model,
single electrons are trapped in the SAW minima and transferred
across the QPC barrier, which is assumed to be long with respect to
the SAW wavelength. As the dot moves towards the center of the
constriction and its size decreases, the Coulomb repulsion between
the trapped electrons restricts their number inside the dot, forcing
some of them to \emph{escape} back to the 2DEG reservoir they where
captured from. Thus, the minimum size of the dot determines the
final number $n$ of transferred electrons.

In this \cite{Shilton1996} and other theoretical models
\cite{Flensberg1999,Robinson2001}, the static barrier of the QPC is
assumed to be long with respect to the SAW wavelength. However, we
believe that this is not entirely valid for our devices. In spite of
large nominal lengths of our QPCs ($\sim2\,\mu$m), their properties
are determined by rather short barriers of around $0.2\,\mu$m
\cite{Gloos2006}, that is about $1/5$ of the SAW wavelength. Such a
short barrier within the QPC channel could reduce the number of
electrons that are further carried in a moving quantum dot.
We emphasize the main difference with respect to the models relying
solely on long QPC barriers
\cite{Shilton1996,Flensberg1999,Robinson2001} where only the
low-energy electrons at the bottom of the dot are transferred across
the constriction. This is no longer the case if the moving quantum
dot approaches a short (though high) barrier. Electrons in the
lowest energy states of the SAW minimum are then held back at the
barrier and return to the reservoir they originated from. Only those
with higher energy are transferred across the QPC and contribute to
the acoustoelectric current.

Those short potential barriers in our nominally long constrictions
could lead to an additional quantization mechanism, independent from
that described by the standard model of 'moving quantum dots'.
Expelling of low-energy electrons at the short static barrier can
take place in series with the quantization mechanisms described in
previous models which rely on the escape or capture of high-energy
electrons \cite{Shilton1996,Flensberg1999,Robinson2001}. The already
quantized current can thus be further reduced at the short barrier,
also to a new quantized level. This could explain
Fig.~\ref{plateaus-2C}: The higher order plateaus (thin
lines/points) are lowered when the second mechanism (thick
line/points) sets in. Thus, this simple model explains the presence
of at least two different kinds of current plateaus. One could
further speculate that, at very narrow channels, more than one
barrier is present in the constriction. One set of acoustoelectric
plateaus could then result from operating the device as a static
quantum dot with two separate barriers of comparable magnitude, as
found in Ref.~\cite{Fletcher2003}.

\begin{acknowledgements}
This work was supported by the European Commission FET Project
SAWPHOTON. P.U. acknowledges support from EC FP6 funding (contract
no. FP6-2004-IST-003673).
\end{acknowledgements}

\end{document}